\documentclass[english,aps,prper,reprint,showpacs,titlepage,longbibliography]{revtex4-1} 

\usepackage[T1]{fontenc}	
\usepackage[latin9]{inputenc}	
\usepackage{geometry}		
\usepackage{graphicx}
\usepackage[above,below]{placeins}	
\usepackage{times}

\usepackage{hyperref} 
\hypersetup{colorlinks=true,urlcolor=blue,citecolor=blue,linkcolor=blue} 
\usepackage{enumitem} 
\setlist{nosep} 

\usepackage[usenames,dvipsnames]{xcolor}

\newcommand{\lr}[1]{#1}
\newcommand{\drdf}[1]{#1}
\newcommand{\hjl}[1]{#1}

\newcommand{\Choice}{{\sc Choice}}
\newcommand{\Execution}{{\sc Execution}}
\newcommand{\Synthesis}{{\sc Synthesis}}
\newcommand{\Contributions}{{\sc Contributions}}
\newcommand{\Understanding}{{\sc New understanding}}
\newcommand{\Emotions}{{\sc Emotional responses}}

\newcommand{\choice}{{\sc choice}}
\newcommand{\execution}{{\sc execution}}
\newcommand{\synthesis}{{\sc synthesis}}
\newcommand{\contributions}{{\sc contributions}}
\newcommand{\understanding}{{\sc new understanding}}
\newcommand{\emotions}{{\sc emotional responses}}

\newcommand{\addquote}[2]{
\begin{itemize}
\item[] ``#1''
\end{itemize}
}

\newcommand{\demotable}{
\begin{table}
\caption{\label{tab:demo}Demographics of physics bachelor's degree recipients at the five universities in our study. Data were averaged over five years for Universities A through D, and ten for E. Numerical table entries in the bottommost row represent average number of physics bachelor's degrees awarded per year. All other numerical entries are percentages. Data were provided by our partner institutions.}
\begin{ruledtabular}
\begin{tabular}{l*{5}{c}}
&\multicolumn{5}{c}{University} \\ \cline{2-6}
Demographic group &A &B &C &D &E \\ \hline
Women\footnotemark[1] &12 &34 &23 &19 &18 \\
Men\footnotemark[1] &88 &66 &77 &81 &82 \\
Native American or Native Alaskan & 0 & 0 & 0 & 0 & 1 \\
Native Hawaiian or Pacific Islander & 0 & 0 & 7& 2 & 0 \\
Black or African American & 0 & 0 & 7 & 3 & 1 \\
Asian American & 1& 5& 17 & 0& 6 \\
Latina, Latino, or Hispanic & 0 & 2 & 33 & 52 & 4 \\
White &87 &73 &23 &41&79 \\
Other or multiple races or ethnicities & 6 & 5 & 10 & 0 & 0 \\
Unknown race or ethnicity & 7 & 2 & 3 & 0 & 7 \\
Temporary visa holders\footnotemark[2] & 0& 13 & {}\footnotemark[3] & 2 & 2 \\ \hline
Physics bachelor's degrees per year & 20 & 25 &6 & 12 & 47 \\
\end{tabular}
\end{ruledtabular}
\footnotetext[1]{Partner institutions reported only binary gender categories.}
\footnotetext[2]{Partner institutions did not provide information about the race or ethnicity of temporary visa holders.}
\footnotetext[3]{Data were unavailable.}
\end{table}
}

\newcommand{\model}{
\begin{figure}\center
\includegraphics[width=0.85\columnwidth]{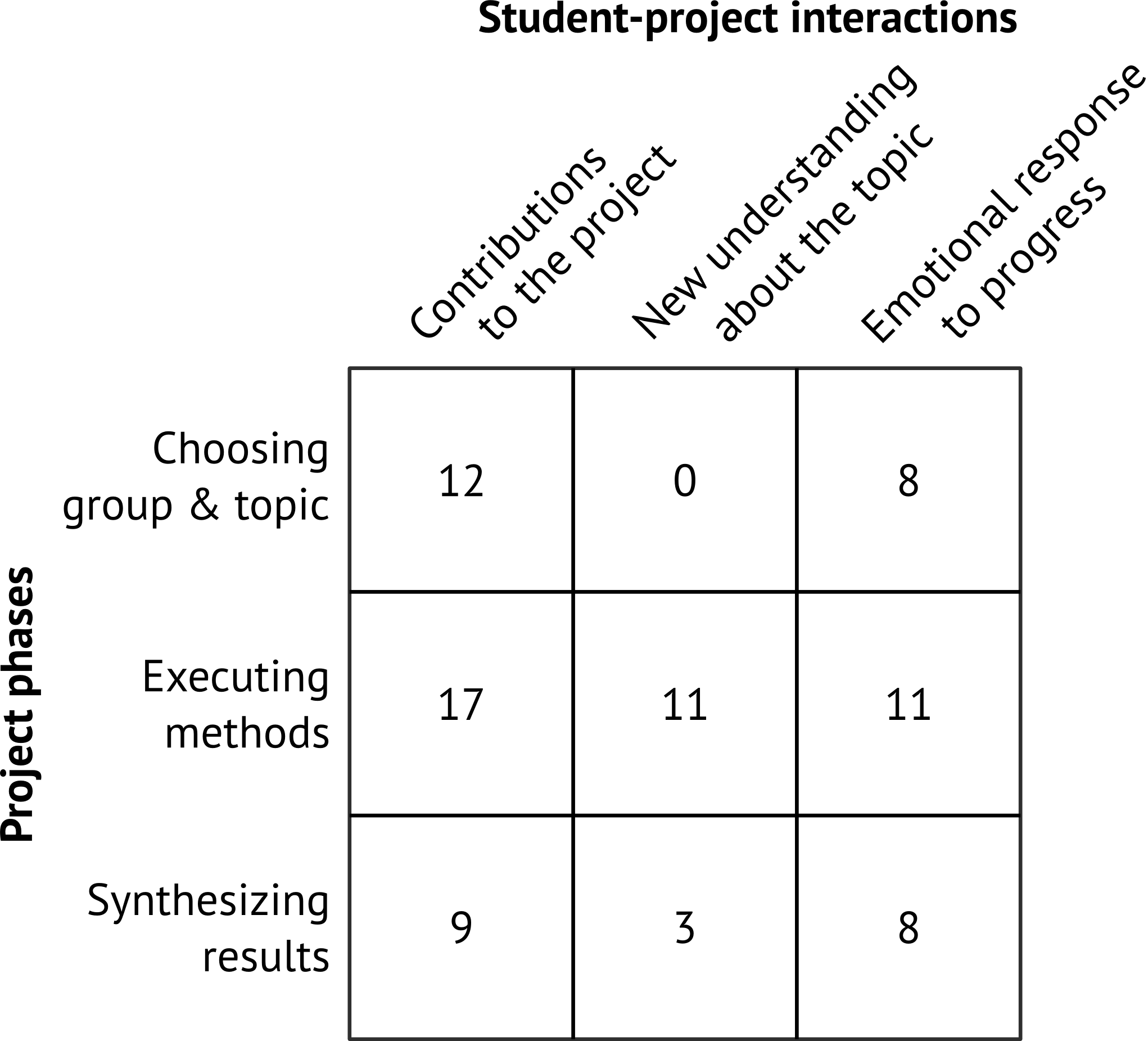}
\caption{\label{fig:model}Preliminary model. Our model conceptualizes ownership as a relationship between student and project that develops over time. We visualize ownership as a grid whose rows and columns correspond to project phases and student-project interactions, respectively. The numbers in each grid entry represent the number of participants whose conceptions of ownership included features that align with the corresponding row and column headers.}
\end{figure}
}

\begin{document}

\begin{titlepage}

\title{Preliminary model for \drdf{student ownership of projects}}

\author{Dimitri R. Dounas-Frazer}
\affiliation{Department of Physics and Astronomy, Western Washington University, Bellingham, WA, USA 98225}
\affiliation{Department of Physics, University of Colorado Boulder, Boulder, CO, USA 80309}
\affiliation{JILA, National Institute of Standards and Technology and University of Colorado Boulder, Boulder, CO, USA 80309}

\author{Laura R\'ios}
\affiliation{Department of Physics, California Polytechnic State University, San Luis Obispo, CA, USA 93407}
\affiliation{Department of Physics, University of Colorado Boulder, Boulder, CO, USA 80309}
\affiliation{JILA, National Institute of Standards and Technology and University of Colorado Boulder, Boulder, CO, USA 80309}

\author{H. J. Lewandowski}
\affiliation{Department of Physics, University of Colorado Boulder, Boulder, CO, USA 80309}
\affiliation{JILA, National Institute of Standards and Technology and University of Colorado Boulder, Boulder, CO, USA 80309}


\begin{abstract}
In many upper-division lab courses, instructors implement multiweek student-led projects. During such projects, students may design and carry out experiments, collect and analyze data, document and report their findings, and collaborate closely with peers and mentors. To better understand cognitive, social, and affective aspects of projects, \lr{we conducted an exploratory investigation of student ownership of projects}. Ownership is a complex construct that refers to, e.g., students' willingness and ability to make strategic decisions about their project. Using data collected through surveys and interviews with students and instructors \lr{at five institutions}, we developed a preliminary model for \drdf{student ownership of projects}. Our model describes student \hjl{interactions with} the project during three phases: choice of topic, execution of experiment, and synthesis of results. Herein, we explicate our model and demonstrate that it maps well onto students' and instructors' conceptions of ownership and \hjl{ideas presented in} prior literature.\clearpage
\end{abstract}

\maketitle
\end{titlepage}

\section{Introduction and background}

The study and transformation of undergraduate \drdf{lab} courses is a \drdf{priority} for the physics education community~\cite{AAPT2014,DBER2012}. Our team is particularly interested in studying and supporting lab courses in which students have the opportunity to carry out their own experimental physics projects over multiple weeks. To this end, we have been exploring the construct of student ownership of projects~\cite{Stanley2016,Dounas-Frazer2017,Dounas-Frazer2018a}. Here, we present work in progress toward developing a model for student ownership of projects. Such a model could facilitate future studies of ownership in lab courses and possibly other settings. It may also be a useful tool for instructors and students alike to reflect on their shared experiences during lab projects and make informed decisions about changes to the learning environment.

We build \drdf{on} existing conceptions of ownership, \lr{including Wiley's (2009) review of relevant literature. He} inferred that student ownership typically refers to students' control over and responsibility for their education, their commitment to participate in their education, or their \drdf{personal connection to} some aspect of their education~\cite{Wiley2009}.
In the context of students working on projects in undergraduate physics and biology courses, student ownership has been linked to motivation, interest, and a sense of pride in overcoming challenges after an extended period of struggle~\cite{Milner-Bolotin2001,Hanauer2014,Little2015}. Ownership has also been described as an individual and group phenomenon~\cite{Enghag2004,Enghag2008,Enghag2009}, an interaction between student and leaning environment~\cite{Hanauer2012}, and a property that develops over time~\cite{Milner-Bolotin2001,Enghag2004}.

\lr{To the best of our knowledge, the most recent diagrammatic models for student ownership of physics projects are at least 15 years old~\cite{Milner-Bolotin2001,Enghag2004}.} Milner-Bolotin (2001) used a Venn diagram to portray ownership as the intersection of three components of learning: finding personal value, taking responsibility, and feeling in control~\cite{Milner-Bolotin2001}. Enghag (2004) used a flow chart to show that ownership, motivation, and competence can be mutually informative phenomena~\cite{Enghag2004}. Last, Savery (1996) depicted ownership as consisting of four psychological quadrants: metacognitive and cognitive factors (e.g., constructing knowledge), affective factors (e.g., motivation), personal and social factors (e.g., teamwork skills), and individual factors (e.g., attitude toward learning)~\cite{Savery1996}. \lr{In our model development efforts, we build from these and other prior models for ownership, and we take into account results from recent and ongoing research on student ownership in physics~\cite{Stanley2016,Dounas-Frazer2017,Dounas-Frazer2018a} and other contexts (e.g., Refs.~\cite{Wiley2009,Hanauer2014,Hanauer2012}). Doing so is an important part of developing a theoretically generalizable model~\cite{Eisenhart2009}.}

Previously, we have used surveys, interviews, and other sources of information to study student ownership of projects in upper-division physics labs~\cite{Stanley2016,Dounas-Frazer2017,Dounas-Frazer2018a}. Our prior work added nuance to earlier ideas about ownership. For example, we found that, in addition to positive feelings like happiness or joy~\cite{Hanauer2014}, students can experience a wide range of fluctuating emotions that contribute to their sense of ownership, including frustration, tedium, and overwhelmedness~\cite{Dounas-Frazer2017,Stanley2016}. We also found that students' interest in a project and their sense of project ownership can sometimes develop in tandem~\cite{Dounas-Frazer2017}, complicating the notion that initial interest in a project is a necessary precursor to ownership~\cite{Hanauer2012,Milner-Bolotin2001}.  More recently, in alignment with Savery's (1996) model of ownership as consisting of individual beliefs~\cite{Savery1996}, we found a moderate positive correlation between students' views about the nature of experimental physics and their sense of project ownership~\cite{Dounas-Frazer2018a}.

\hjl{In this paper,} we describe a qualitative study in which we interviewed instructors and students from five \hjl{colleges or} universities about their conceptions of student ownership. Based on interview responses and prior literature, we propose a preliminary model for student ownership of \hjl{lab} projects.

\section{Research contexts}

In this section, we describe our criteria for choosing sites, and we provide demographic and other contextual information about each site. This information is needed to establish and constrain the generalizability of our results~\cite{Eisenhart2009}. Moreover, providing demographic and contextual information facilitates metastudies of the physics education literature (e.g., Ref.~\cite{Kanim2017}), and it disrupts the erroneous subconscious assumption that physicists and physics students are white men unless otherwise specified (cf. Ref.~\cite{Parks2012}).

\lr{When selecting sites, we sought partnerships with a small number of instructors with whom we had existing professional relationships through our participation in communities dedicated to physics labs instruction (e.g., the Advanced Laboratory Physics Association). Additionally,} we wanted to avoid partnering only with instructors who taught at selective Predominantly White Institutions (PWIs) because those institution types are overrepresented in the physics education research literature~\cite{Kanim2017}.
Ultimately, we partnered with instructors at five institutions in the Western or Midwestern United States: one selective Baccalaureate College, two inclusive Master's Colleges or Universities, and two selective Doctoral Universities. Two institutions were Hispanic-Serving Institutions (HSIs), and three were PWIs. Each instructor taught an upper-division physics lab course with a total enrollment of 12 to 24 students. The course was required for at least one physics bachelor's degree track, and most or all of the enrolled students were physics majors. As a result, the enrolled students were demographically similar to physics bachelor's degree recipients at the same institution. A summary of degree recipient demographics is provided in Table~\ref{tab:demo}. 

Each course in our study included a project component during which students worked in groups of two to four. Project topics included constructing a chaotic pendulum, measuring the absorption spectrum of a rubidium vapor, and achieving thermal lensing in soy sauce. Projects ranged in duration from four to seven weeks. Previous work in undergraduate physics and biology labs  suggests that projects of this length are sufficient for students to develop a sense of ownership~\cite{Stanley2016,Dounas-Frazer2017,Hanauer2018}, even if students complete only some of their initial goals for the project~\cite{Dounas-Frazer2017,Gin2018}.

\demotable

\section{Data collection and analysis}

Overall, 87 students agreed to participate in one or more aspects of our study, corresponding to a student participation rate of about 95\%. We administered weekly open-ended reflections throughout the project portion of the course. Reflection prompts were similar to those we developed and used in a previous study~\cite{Dounas-Frazer2017}, and they asked about students' goals, challenges, and successes while working on their projects. In total, we collected over 350 reflections.

We conducted post-project interviews with 4 instructors and 15 students, corresponding to a student interview participation rate of about 16\%. The instructor from University A did not participate in an interview; neither did students from University E. Possible explanations will be discussed in forthcoming work. All interviewees were provided with monetary incentives. Ten interviewees self-identified as white men and five as white women. One self-identified as a mixed-race Central American Latina woman, one as a white and Japanese woman, one as an Asian and Middle Eastern woman, and one as an ethnically Chinese man from Hong Kong. 

Instructor interviews lasted 80 to 110 minutes, for a total of 6 hours. They were semi-structured and focused primarily on participants' strategies for\hjl{, and experiences with,} implementing projects in their lab courses. During interviews with students, the interviewer and interviewee collaboratively filled out a life grid. \hjl{By ``life grid,'' we mean a digital spreadsheet whose rows correspond to intervals in time and whose columns represented different aspects of the project, such as revisions to apparatus or interactions with peers~\cite{Blane1996,Abbas2013}}. Further details about this approach are provided in Ref.~\cite{Rowland2019}. Student interviews lasted 55 to 80 minutes, for a total of 17 hours. In interviews with instructors and students alike, the interviewer asked participants to define student ownership and to provide examples from their own experiences. The first and second authors transcribed recordings, and the transcripts are the \hjl{data we} analyzed. We limit our discussion to instructor and student responses to interview questions about their conceptions of ownership.

Our analysis process consisted of two rounds. During each round, the first and second authors read through transcripts, identified excerpts that corresponded to particular ideas about what ownership is or how it can be fostered, and discussed their rationales and interpretations. We referenced existing literature about student ownership and reflected on our own experiences working on projects. At the end of each round, the two authors reached consensus on a tentative model for student ownership of projects, and they collaboratively coded the transcripts according to those models. Coding entailed grouping printouts of transcript excerpts into piles according to the major features of the model and using qualitative data analysis software to keep digital records of those groupings. When grouping excerpts and assigning codes, transcribed utterances that represented a complete idea were treated as a single reference. After coding, the whole research team discussed elements of the model and representative transcript excerpts.
Throughout this process, we generated a variety of visual representations to guide our thinking. During the second round, we settled on the grid similar shown in Fig.~\ref{fig:model}.

In Fig.~\ref{fig:model}, rows represent phases of the project: choosing the project topic, carrying out the research, and identifying and reporting outcomes (cf. Ref.~\cite{Enghag2004}). Columns represent interactions between students and projects: making intellectual or material contributions to the project, developing new understanding about the project topic, and experiencing emotional responses to the group's progress (or lack thereof) on the project. When using Fig.~\ref{fig:model} as an analysis scheme to code transcripts, each excerpt was assigned one of nine groups depending on project phase and type of interaction. Fig.~\ref{fig:model} constitutes our preliminary framework for student ownership of projects, and its development was the major output of the second round of our analysis.

\section{Preliminary model}


We interpret student ownership of a project as a type of relationship between student and project. By ``project,'' we mean the topic, apparatus, and methods for doing \hjl{research,} as well as the people with whom the research is done. Our interpretation is similar to those of Milner-Bolotin (2001) and Hanauer et al.\ (2012) in which ownership was framed as an interaction between student and the educational environment or learning process~\cite{Milner-Bolotin2001,Hanauer2012}. Like other relationships, ownership evolves in time~\cite{Milner-Bolotin2001} and is characterized by a set of interactions between the things that are related, namely, students and projects. Thus, the types of student-project interactions typical of ownership-style relationships depend on the project phase, as implied by Fig.~\ref{fig:model}. 

Project phases are defined as follows.
\Choice\ refers to the phase in which students choose their project topic and group members. Choice of topic can be constrained by group members or instructors according to appropriateness with respect to learning goals, resources, time constraints, or collective team interest. Choice of group can also be constrained by instructors' assignments.
\Execution\ refers to the phase in which students execute their research by designing, building, or troubleshooting apparatus; writing, using, or debugging computer code; or collecting, interpreting, or analyzing data. Student execution of research can be guided by instructor mentorship or team collaboration.
\Synthesis\ refers to the phase in which students synthesize their notes, results, and new knowledge to create summative reports and presentations. Reports and presentations may have an audience that extends beyond instructors and students in the class.

Categories of student-project interaction are defined as follows.
\Contributions\ refers to a student's own intellectual or material contributions to the project, potentially in partnership with peers and mentors. These contributions change or advance the project in some way.
\Understanding\ refers to a student's own knowledge about, or understanding of, the project or topic, and how the project is changing or advancing that knowledge or understanding.
\Emotions\ refers to a student's emotive responses to the project, their own knowledge, or lack thereof; it is how a student feels about what they or others are doing and knowing.

The model for ownership represented by Fig.~\ref{fig:model} has precedent in the literature. For example, Enghag (2004) defined similar project phases~\cite{Enghag2004}. Similarly, all three categories of interaction align with others' conceptions of ownership. For example, when students make their own contributions, they are demonstrating a type of agency that is often associated with ownership~\cite{Wiley2009,Milner-Bolotin2001} and which is constrained by input from peers and mentors~\cite{Enghag2008,Hanauer2012}. Enghag (2004) and Savery (1996) noted connections between students' sense of ownership and their competence with or knowledge about the project~\cite{Enghag2004,Savery1996}, and many researchers have described the affective components of ownership (e.g., Refs.~\cite{Dounas-Frazer2017,Hanauer2014,Savery1996}).

\section{Mapping to participants' conceptions}

\model

\lr{When using Fig.~\ref{fig:model} as a coding scheme, each transcript received multiple codes. Each interaction category was identified in a majority of transcripts, and likewise for each project phase. This indicates a good mapping between the model and the data.} In this section, we elaborate on this mapping using excerpts from interviews with Mac (an instructor) and five students: Olivia, Lance, Jordyn, Gandalf, and Heather. All names are pseudonyms.

\Choice\ was co-coded with \contributions\ or \emotions, as evidenced by Olivia and Mac. Olivia described how her group's search for a ``final project-worthy'' topic (i.e., intellectual \contributions\ to the project) supported them to feel ownership of the project:
\addquote{\ldots\ [W]e actually chose ours. We had to go out and find, `Okay, here's a paper that looks interesting. Is it something we could do? Is is something that is kind of final project-worthy?' And so we had to go out and find it. Like, go out and find something we could do for our final project. 'Cause they [the instructors] had given us some suggestions, like redo Millikan's oil drop, speed of light stuff, and calculating things. Things we already knew. But this one, it required us to, we found it ourselves. And so that was kind of our own little ownership.}{Olivia, student}
Mac reasoned that \emotions\ to project choice are ``a piece of ownership.'' Specifically, she described investment, enthusiasm, and commitment:
\addquote{I think that's a piece of ownership, too, is level of investment. \ldots\ As people are trying to decide on what their projects are, maybe two members are really enthusiastic about this particular project and the third one is not really, but will defer to these two in deciding what the topic is going to be, or whatever. So then, they aren't necessarily maybe having as big of a role in deciding the topics, but they nonetheless still feel committed to making this happen.}{Mac, instructor}
When describing the process of choosing a project, both Olivia and Mac referenced input from or negotiations with peers and mentors.

There were no instances where \choice\ was co-coded with \understanding. However, we believe that new understanding can indeed be generated during the first phase of a project. Olivia hints at a connection between project choice and generating new knowledge by framing projects about ``[t]hings we already knew'' in contrast to the project she and her team decided to pursue. Further, Mac requires her students to write project proposals during the \choice\ phase, and it is likely that students' understanding of the project topic is advanced during that assignment.

\Execution\ was co-coded with all three interaction categories, as evidenced by Lance and Jordyn. When recalling a time when he felt ownership of a project, Lance \drdf{described his} intellectual \contributions\ to the troubleshooting process:
\addquote{I remember one day, a few weeks into the project, we were troubleshooting something, and it wouldn't work. And he [a professor] was suggesting this thing that I kept trying, and it wasn't working. So he came over for a few minutes, and then I kind of realized, `Oh, well your mentor doesn't always know what they're doing, and you're not just there to perform ideas. You should also contribute.' So I started to really think through that project, and I ended up realizing why we were having this problem, and then I was able to fix it. So that was kind of this idea of, `Oh, when I'm working on a project, I really need to be contributing ideas, not just performing the ideas of my mentor or the professor.'}{Lance, student}
Jordyn described connections between ownership and development of \understanding\ while working on the project. In terms of her \emotions, she framed this process as simultaneously uncomfortable and interesting:
\addquote{\ldots\ [P]art of ownership is sort of that uncomfortable realizing that you have something of a monopoly of understanding, or at least a depth of understanding that isn't had by other people. Or, at least, obviously you should not be monopolizing the understanding from your group mates, but you understand your project supposedly better than your professors, and it sort of maybe that uncomfortable feeling. Uncomfortable, but also sort of interesting feeling that defines ownership of a project.}{Jordyn, student}
While both Lance and Jordyn described intellectual contributions to the projects, other participants described material contributions, especially investment of time.

\Synthesis\ was also co-coded with all three interaction types, as evidenced by Gandalf and Heather. Gandalf reasoned that he had ownership of his project because, during the final phase of the project, he was uniquely \drdf{committed} to investing the time and effort required to get ``\drdf{really} good images'' for his group's summative report and presentation:
\addquote{I guess I kind of felt like I had a lot of ownership in this project because, at that point [the final week of the project], I kind of felt like I was, like, clearly the most motivated person to get really good images with our final goals. So I felt a lotta ownership there, just 'cause I was putting in the most time. Um, yeah, and then I had, along with some other people, had gotten these good results. So I felt good about that, and I felt like they were mostly due to my effort at that point.}{Gandalf, student}
In addition to connecting his \contributions\ to his sense of ownership, Gandalf also described his \emotions\ during the final phase of the project: he was motivated to get good images, and he felt good about the images he acquired. Heather said she felt ownership of her project, and she described how the poster session in particular helped her recognize the \understanding\ she gained by completing the project:
\addquote{I think, slowly over the course of doing the project, I've been learning all these things, but never to the extent that I could explain it to someone else and have them understand it, too. And I guess it didn't have to be the poster session, but that was just the first opportunity where I saw it. \ldots\ People understood things that I was saying as information that I had to teach myself and then I could pass on to other people.}{Heather, student}
Like Gandalf and Jordyn, many other students also drew connections between their sense of ownership and their interactions with projects during the \synthesis\ phase. However, no instructors drew similar connections; instructors' conceptions of student ownership focused only on student-project interactions during \choice\ and \execution.

\section{Future directions}
We have developed a preliminary model for student ownership of projects in physics lab courses. Our model frames ownership as a relationship between student and project that evolves over time and is characterized by three categories of student-project interactions. To develop this model, we relied on previous literature and interviews with 4 instructors and 15 students from 5 institutions. Our model is preliminary in the sense that we have not yet mapped it to our full data set. The full data set includes in-depth descriptions of project experiences from both instructor and student perspectives.

In ongoing work, we are mapping the model onto the full data set. Doing so will allow us to add nuance and specificity to our conception of ownership. For example, we anticipate identifying instances where students generate new understanding about the project topic during the process of choosing which project to work on. Ultimately, this work will culminate in a model that can be used to facilitate the study and design of lab courses in which students pursue their own multiweek research projects.

\acknowledgments{The authors acknowledge Benjamin Pollard for useful conversations about data interpretation. This material is based upon work supported by the NSF under Grant Nos.
DUE-1726045, and 
PHY-1734006. 
}

\bibliography{./perc2019_database} 

\end{document}